\documentclass[12pt]{iopart}

\usepackage{graphicx}
\begin{document}

\title[Moment switching in nanotube magnetic force probes]{Moment switching in nanotube magnetic force probes}

\author{John R Kirtley$^{1,2,3}$,
        Zhifeng Deng$^4$,
        Lan Luan$^4$,
        Erhan Yenilmez$^1$,
        Hongjie Dai$^5$,
        and Kathryn A Moler$^{1,4}$}

\address{
$^1$ Department of Applied Physics and Geballe Laboratory for Advanced Materials,
Stanford University, Stanford, California 94305 USA}
\ead{jkirtley@stanford.edu}
\address{
$^2$ IBM Watson Research Center, Route 134 Yorktown Heights, NY 10598 USA}
\address{
$^3$ Faculty of Science and Technology and MESA$^+$ Institute for Nanotechnology,
University of Twente, P.O. Box 217, 7500 AE Enschede, The Netherlands}
\address{
$^4$ Department of Physics and Geballe Laboratory for Advanced Materials,                 %
Stanford University, Stanford, California 94305 USA}
\address{
$^5$ Department of Chemistry, Stanford University, Stanford, California 94305 USA}

\begin{abstract}

A recent advance in improving the spatial resolution of magnetic force microscopy (MFM)
uses as sensor tips carbon nanotubes
grown at the apex of conventional silicon cantilever pyramids and coated with a thin
ferromagnetic layer \cite{deng2004}.
Magnetic images of high density vertically recorded media using these tips
exhibit a doubling of the spatial frequency under some conditions \cite{deng2004}.
Here we demonstrate that this
spatial frequency doubling is due to the switching of the moment direction of
the nanotube tip. This results in a signal which is proportional to the absolute
value of the signal normally observed
in MFM. Our modeling indicates that a significant fraction of the tip volume
is involved in the observed switching, and that it should be
possible to image very high bit densities
with nanotube magnetic force sensors.

\end{abstract}

\begin{figure}
\includegraphics[width=6.0in]{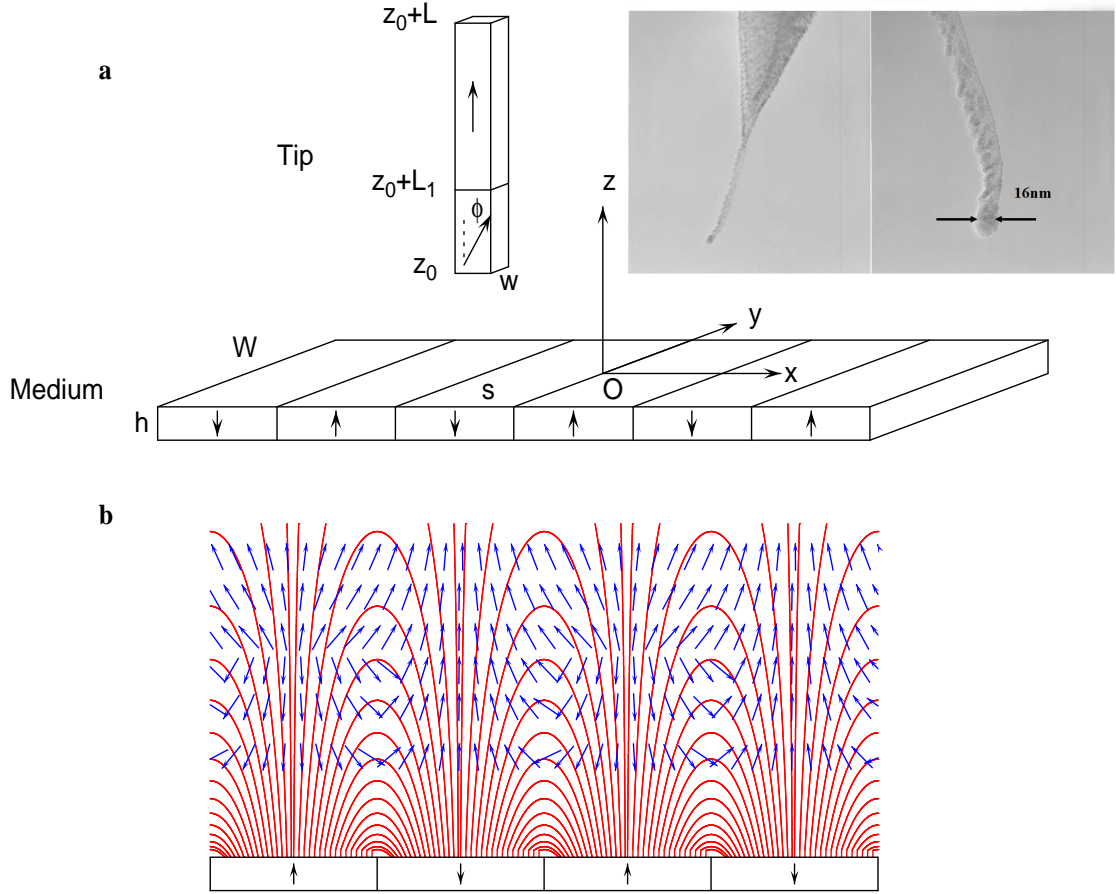}
\caption{\label{fig:tipandgeo}
(a) Tip and sample geometry in model. The inset is a
scanning electron microscope image of the cobalt-coated nanotube tip.
(b) Calculated field lines for a 300 kfci track
above the center of the domains in the
magnetic medium, using the parameters
described in the text. The arrows indicate calculated moment orientations
for a series of tip-sample spacings of 15, 22.5, 28, 33, 39, and 45 nm using
the model and parameters described in the text. At small tip-sample
spacings the tip moment flips to be always anti-aligned with the moment
directly below the tip. At large tip-sample spacings the tip moments are only
slightly perturbed by the medium fields.
}
\end{figure}

Spatial period doubling has been observed for several carbon nanotube tips with
different track widths and bit densities.
The MFM images reported here were made using the nanotube tip shown in the inset
of Figure \ref{fig:tipandgeo}. It was approximately 250nm long, with a ferromagnetic coating
to a total tip diameter of 16nm.
The means of producing metal-coated carbon nanotube tips on AFM cantilevers
and the techniques for imaging magnetic
media using these tips have been described previously \cite{deng2004,deng2006}.
Briefly, carbon nanotubes were
grown using wafer-scale chemical vapor deposition
at the apexes of the pyramids of commercial silicon tips intended for tapping mode atomic
force microscopy.
For the present measurements the nanotubes were shortened to a length of
about 250 nm using an electrical cutting method \cite{yenilmez2002}, aligned approximately perpendicular
to the cantilever using a focused ion beam \cite{deng2006},
and then coated to a total thickness of about 16 nm
with a Ti/Co/Ti trilayer by e-beam evaporation from a direction parallel
to the nanotube long axis
(see the inset in Figure \ref{fig:tipandgeo}). The magnetic imaging was
done using the Tapping/Lift$^{TM}$ mode of a Digital Instruments Nanoscope III SPM at room
temperature in air. In this mode, the topography of the sample is first determined for each line
with a scan at low tip-sample spacing $z_0$, then the tip is retracted a specified distance and
a second line scan is made while recording the deviation of the phase angle $\delta$
of the cantilever
response with the cantilever driven slightly below its resonance frequency. $\delta$ is
proportional to $dF_z/dz_0$, the derivative of the force
on the cantilever
with respect to $z_0$.

\begin{figure}
\includegraphics[width=6.0in]{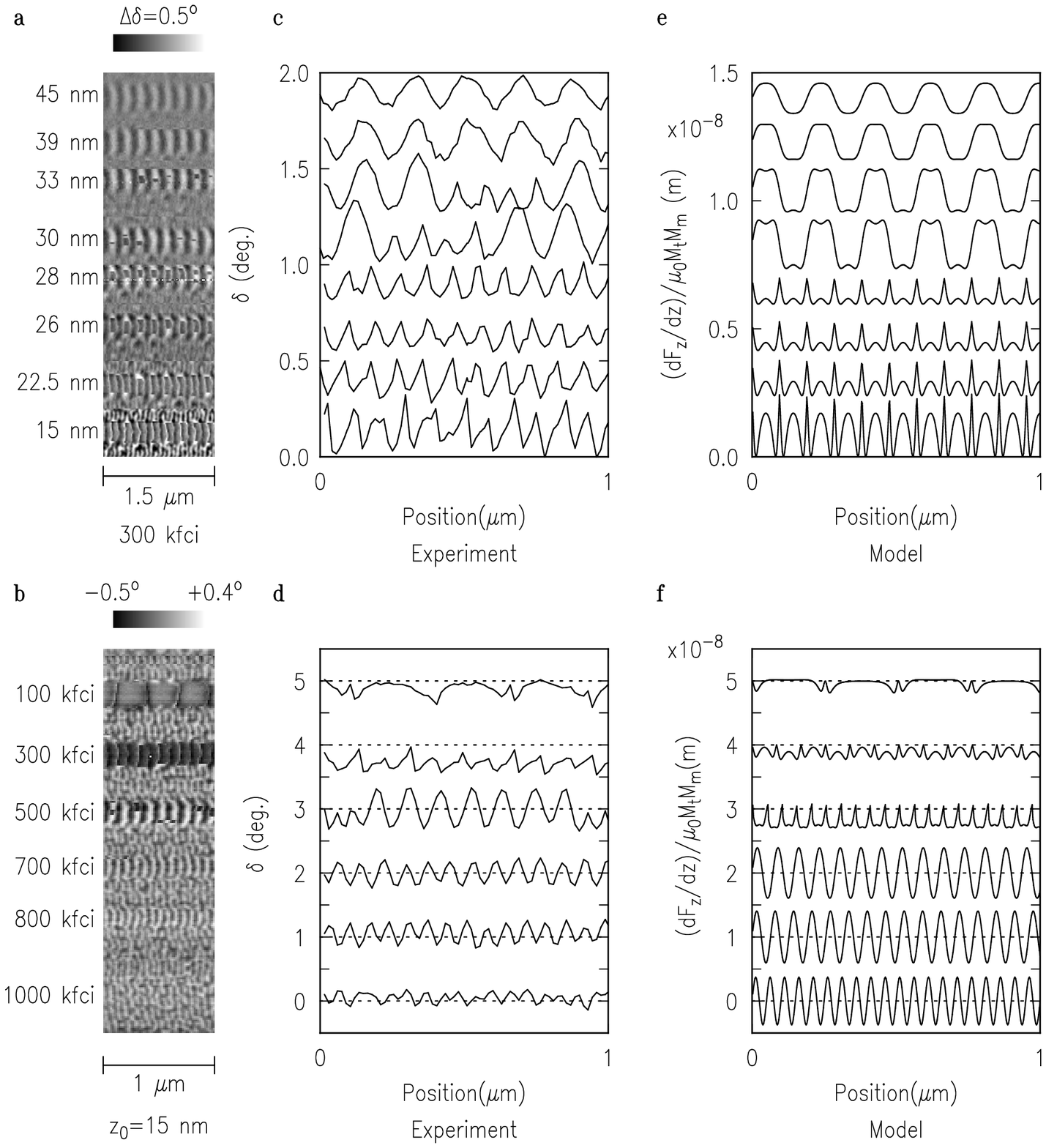}
\caption{\label{fig:natnt2}
(a) Collection of phase shift images for a 300 kfci track of alternating vertically recorded media
moments at different lift heights. Each line of the image has an offset
such that the average phase shift value for that line is zero.
(b) Image of a section of vertically recorded media with a series of tracks with different bit
densities, at a lift height of 15 nm. A background has been subtracted from this image such that the
regions between the tracks have zero averaged phase shifts.
(c) Cross-sections through the centers of the tracks in (a).
(d) Cross-sections through the centers of the
tracks in (b). The zero phase shift levels for each cross-section are indicated by dashed lines.
(e) Modeling of (c) as described in the text. The lines in (c) and (e) have been offset vertically for clarity.
(f) Modeling
of the tracks in (d) as described in the text, with the zero phase shift levels indicated by
dashed lines.}
\end{figure}

The images presented here were made on a vertically polarized magnetic medium.
A collection of phase shift images at 300 kilo-flux changes per inch (kfci)
and different tip-sample spacings ($z_0$'s)
are shown in Figure \ref{fig:natnt2}(a), and at constant lift height (15nm) and different bit densities
in Figure \ref{fig:natnt2}(b). Cross-sections of the data through the center of the tracks are displayed
in Figure \ref{fig:natnt2}(c,d). Modeling of this data as described below is shown
in Figure \ref{fig:natnt2}(e,f).
At high values of $z_0$ and high bit densities the phase images
have the same period as written, but at low $z_0$'s and low bit densities anomalies in the images
gradually develop into sharp features with double the original periodicity. These
anomalies are due to switching of the orientation of the magnetic moment of the tip.
This can be demonstrated most convincingly by inspecting the images and cross-sections of
Figure \ref{fig:natnt2}(b,d). In this case a background
from the sections of the image without written bits
has been subtracted out, and it can be seen that at low bit densities the phase shift always
stays below the average background level, keeping the tip-sample force attractive
by switching of the tip moment direction just as the force derivative crosses zero.

This conclusion is supported by detailed modeling:
We assume that the magnetic medium is composed of
slabs of length $s$ in the $x$ direction, height $h$ in the $z$ direction, and width $W$ in the
$y$ direction (Figure \ref{fig:tipandgeo}(a)). The slabs have a uniform magnetization with
moment direction alternating between
parallel and anti-parallel to the $z$-axis direction.
The tip is assumed to have a square cross-section
with width $w$ and length $L$, with the tip end a distance $z_0$ from the upper surface of the medium.
The magnetic fields above the sample (displayed as field lines in Figure \ref{fig:tipandgeo}(b))
were calculated both analytically and numerically.
In our numerical modeling
the vector between the individual medium dipole moments [$x_m,y_m,z_m$] and a position
[$x,y,z$] is $\vec{r}=(x-x_m)\hat{x}+(y-y_m)\hat{y}+(z-z_m)\hat{z}$.
Then the $z$ and $x$-component of the field from the individual dipoles is given by
\begin{eqnarray}
B_z(\vec{r}) & = & \frac{\mu_0 m_m}{4\pi r^3} \left (\frac{3(z-z_m)^2}{r^2}-1 \right ) \nonumber \\
B_x(\vec{r}) & = & \frac{3 \mu_0 m_m}{4\pi r^5} (x-x_m)(z-z_m) ,
\label{eq:z_component}
\end{eqnarray}

with $r=| \vec{r} |$.
$B_y$ is given by setting $x \rightarrow y$ and $x_m \rightarrow y_m$ in the second equation.
If we assume that the tip can also be represented by a sum of point dipoles $\vec{m}_{t}=m_{t} \hat{n}$,
where $\hat{n}$ is a unit vector in the tip moment direction,
the force gradient on the tip is given by
\begin{eqnarray}
\frac{dF_z}{dz_0} &= & \sum_{tip} m_{t} \cos(\phi) \frac{d^2B_z(x,y,z)}{dz^2}  \nonumber   \\
                  &= &
\sum_{tip,medium} \frac{3\mu_0 m_m  m_{t} \cos(\phi)}{4\pi} \left [ \frac{3}{r_{tm}^5} - \frac{30 (z-z_m)^2}{r_{tm}^7}
+\frac{35 (z-z_m)^4}{r_{tm}^9} \right ],
\label{eq:force_gradient}
\end{eqnarray}
where $r_{tm}$ is the distance between the individual dipole elements in the tip and medium, and
$\phi$ is the angle between the tip moment direction and the $z$-axis.

For positions near the center of the tracks we obtained analytical expressions for the magnetic fields
by assuming that the magnetically oriented slabs have
infinite width $W$ in the $y$ direction.
If we take the boundary condition
\begin{equation}
H_z(\vec{r}) |_{z=0^+} - H_z(\vec{r}) |_{z=0^-} =
H_z(\vec{r}) |_{z=-h^-} - H_z(\vec{r}) |_{z=-h^+} = \sigma(\vec{r}),
\label{eq:boundary}
\end{equation}
where $\sigma(\vec{r})$ is the surface magnetic charge,
the magnetic field above the sample can be written as \cite{steifel1998}
\begin{equation}
H_z(\vec{r},z) = \frac{1}{4\pi^2} \int_{-\infty}^{\infty} d^2 \vec{k} A_{H,z}^{z,h} e^{-i \vec{k} \cdot \vec{r}},
\label{eq:steifel_h_z}
\end{equation}
where
\begin{equation}
A_{H,z}^{z,h} = e^{-kz}(1-e^{-kh}) \int_{-\infty}^{\infty} d^2 \vec{r} \, \sigma(\vec{r}) e^{i \vec{k} \cdot \vec{r}}.
\label{eq:a_Hz}
\end{equation}
Taking the surface magnetic charge $\sigma(\vec{r})$ to be uniform in $y$,
\begin{eqnarray}
\sigma(\vec{r}) & =  & M   \hspace{0.30in} 2ns < x < (2n+1)s \nonumber  \\
                & =  & -M  \hspace{0.2in} (2n-1)s < x < 2ns,
\label{eq:sigma}
\end{eqnarray}
We find
\begin{eqnarray}
H_z & = & \frac{M}{2\pi} \sum_{n=-\infty}^{\infty} \left [ \tan^{-1} \frac{z}{(2n+1)s+x} + \tan^{-1} \frac{z}{(2n-1)s+x}
-2 \tan^{-1} \frac{z}{2ns+x} \right . \nonumber \\
    & - & \left . \tan^{-1} \frac{z+h}{(2n+1)s+x} - \tan^{-1} \frac{z+h}{(2n-1)s+x} + 2 \tan^{-1} \frac{z+h}{2ns+x} \right ]
\label{eq:steifel_hz_sum}
\end{eqnarray}
Since $H_z = d\Phi/dz$, $\Phi$ a scalar potential, the $x$-component of the field can be written as
$H_x = d\Phi/dx$, which leads to
\begin{eqnarray}
H_x & = - & \frac{M}{4\pi} \sum_{n=-\infty}^{\infty} \left [ \log \left (\frac{z^2}{((2n+1)s+x)^2}+1 \right )
                                                           + \log \left (\frac{z^2}{((2n-1)s+x)^2}+1 \right )          \right . \nonumber \\
    &   & \left .                                                     -2 \log \left (\frac{z^2}{(2ns+x)^2}+1 \right )
                                                         - \log \left (\frac{(z+h)^2}{((2n+1)s+x)^2}+1 \right )         \right . \nonumber \\
    &   & \left .                                                      - \log \left (\frac{(z+h)^2}{((2n-1)s+x)^2}+1 \right )
                                                         + 2 \log \left (\frac{(z+h)^2}{(2ns+x)^2}+1 \right ) \right ]
\label{eq:steifel_hx_sum}
\end{eqnarray}
$H_y$ is zero by symmetry.
\begin{figure}
\includegraphics[ width=2.0in]{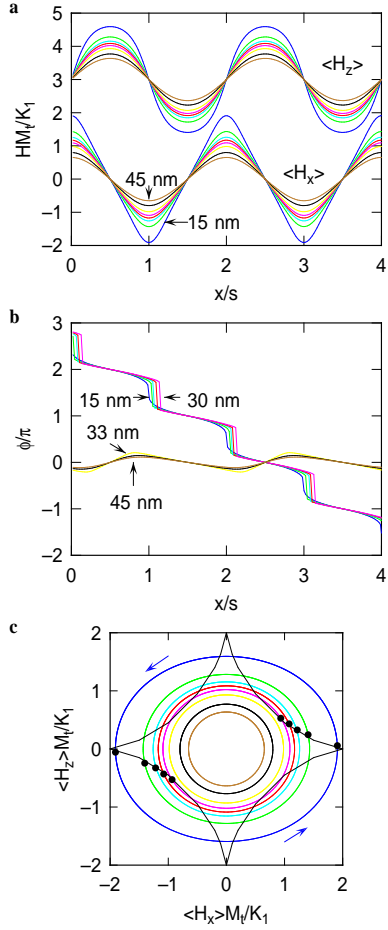}
\caption{
\label{fig:zhifield} (color online)
(a) Calculated magnetic fields, averaged over the tip volume and
multiplied by the tip saturation magnetization $M_t$
divided by the anisotropy parameter $K_1$,
above a vertically recorded magnetic medium with infinite extent in the $y$-direction,
12 nm thick in the $z$-direction, with magnetization
oriented in the $z$-direction and alternating in the $x$-direction with period $2s=185$ nm,
for spacings between the bottom of the tip
and the medium of $z_0$ = 15, 22.5, 26, 28, 30, 33, 39, and 45 nm. The tip has an assumed
square cross-section 16nm on a side, and the averaging is over a tip length $L_1 = 48 nm$.
The calculated magnetic fields $<H_z>$ normal to the magnetic medium
are offset by 3 units. The best fit values $M_tM_m/K_1=17.5$, $\sigma_w/\mu_0L_1K_1=0$
(see main text Figure 3)
were used for these calculations.
(b) Calculated variation of the tip moment orientation angle $\phi$ relative to the $z$ axis.
(c) The ovals show the calculated trajectories of $<H_z>M_t/K_1$ vs
$<H_x>M_t/K_1$ for the various tip heights. The ``asteroid"
plots the values of the critical fields.
The intersections of these two sets
of curves, indicated by solid symbols, are the fields at
which switching occurs in the simulations.}
\end{figure}
Figure \ref{fig:zhifield} displays the calculated fields in the $z$ and $x$ directions (a,c),
the tip moment orientation angle $\phi$ (b),
and the switching fields $<H_z>_c$ and $<H_x>_c$ (c)
for the best fit to the data as described below.
It is interesting to note that the discontinuous switches of $\phi$ are controlled
predominantly by the size of the $x$ component of the field. This can be understood by examining
the trajectory in field that the tip takes in moving from one domain to the next. The ovals in Figure
\ref{fig:zhifield}c are the calculated trajectories of $<H_z>$ vs. $<H_x>$ for the tip heights listed
in the caption. The ``asteroid" is the critical field calculated for the energy functional form
of Eq. 1 of the main text
and has the form first predicted by Stoner and Wohlfarth\cite{stoner1948}.
The tip moment direction is predicted to switch when the field trajectories cross the critical
field asteroid (solid dots in Figure \ref{fig:zhifield}c). This happens for relatively large values of
$|<H_x>|$ and small values of $|<H_z>|$. When the field trajectory crosses the Stoner-Wohlfarth
asteroid at large values of $|<H_z>|$ the tip has already switched to the low energy configuration.

For numerical work the medium was assumed to be composed of a collection of individual
dipole moments $\vec{m}= m_m \hat{z}$, where $m_m = \pm M_m v$, with $M_m$ the saturation magnetization and
$v$ the volume of the individual medium elements. In what follows we take both the tip and medium
volume elements to be cubes 4 nm on a side. This results in agreement to within a few percent between
our numerical work and analytical expressions for the medium magnetic fields.
Halving the size of the volume elements (to cubes 2 nm on a side) changes the calculated
force derivative curve for
$z_0$ = 30 nm (see Figure \ref{fig:natnt2}c) by about 2\%.

To model the dynamics of the tip flip process,
we conceptually divide the tip into
two domains, one with length $L_1$ close to the medium, the other with length $L-L_1$ further away.
Each has sufficiently strong exchange fields that the entire volume within each domain
has the same moment orientation\cite{kittel1946}. The section of the tip furthest from the medium
is assumed to have its
moment parallel to the $z$ axis; that closest to the medium has its moment at an
angle $\phi$ relative to the $z$-axis (Figure \ref{fig:tipandgeo}).
Then the energy of the tip in an external magnetic field can be written as:
\begin{eqnarray}
E(\phi) & = & \mu_0 w^2 L_1 K_1 \left [ \frac{\sigma_w}{\mu_0 L_1 K_1}\frac{1}{2} (1 - \cos \phi)
+\sin ^2 \phi  \right . \nonumber \\
& + & \left . \frac{M_t}{K_1}(<H_z> \cos \phi + <H_x> \sin \phi ) \right ]
\label{eq:phienergy}
\end{eqnarray}
where we have taken the simplest non-trivial forms for the domain wall energy (first term)
and the
anisotropy energy (second term)\cite{stoner1948,bonet1999}. The third term in Eq. \ref{eq:phienergy}
is the energy of the dipole moments of the tip in the external magnetic field. Here $\mu_0K_1$ is
the anisotropy energy density, $\sigma_w$ is the domain wall energy per unit area, $M_t$ is the tip saturation
magnetization, and $<H_z>$ and $<H_x>$ are the magnetic fields in the $z$ and $x$ directions respectively
averaged over the tip volume from
$z=z_0$ to $z=z_0+L_1$.
To simulate the magnetic force images, the tip moment is
at first assumed to be parallel to the $z$-axis. The tip is moved to a new position, the local fields
are calculated and averaged over the tip volume,
$\phi$ is moved to the new local minimum in energy (Eq. \ref{eq:phienergy}), the force gradient
is calculated, and the process is repeated. This modeling results in the cross-sections displayed
in Figure \ref{fig:natnt2}(c,f), which reproduce the
absence of tip switching at high bit densities and high lift heights, and the presence of
tip switching at low bit densities and low lift heights. When tip switching occurs, the modeling also
reproduces the fact that the tip-sample force gradient always stays negative, with the tip moment
reversing as the z-component of the field crosses zero.
The quantitative interpretation of MFM images
in the presence of tip switching is straightforward once it is recognized that
the phase shift $\delta$
is proportional to the negative of the absolute value of the tip sample force gradient (-$ |dF_z/dz_0|$).
The experimental phase shift oscillation amplitudes decrease much more rapidly than the modeling
for bit densities above about 500 kfci
(Figure \ref{fig:natnt2}f). We believe that this is because
the as written bits do not have as abrupt moment orientation reversals as our idealized model.
Our modeling indicates that
nanotube tips with the geometry of Figure \ref{fig:tipandgeo} could be used to image bits with
sharp moment direction transitions with densities
above 2000 kfci (13 nm/flux reversal).

\begin{figure}
\includegraphics[ width=4.0in]{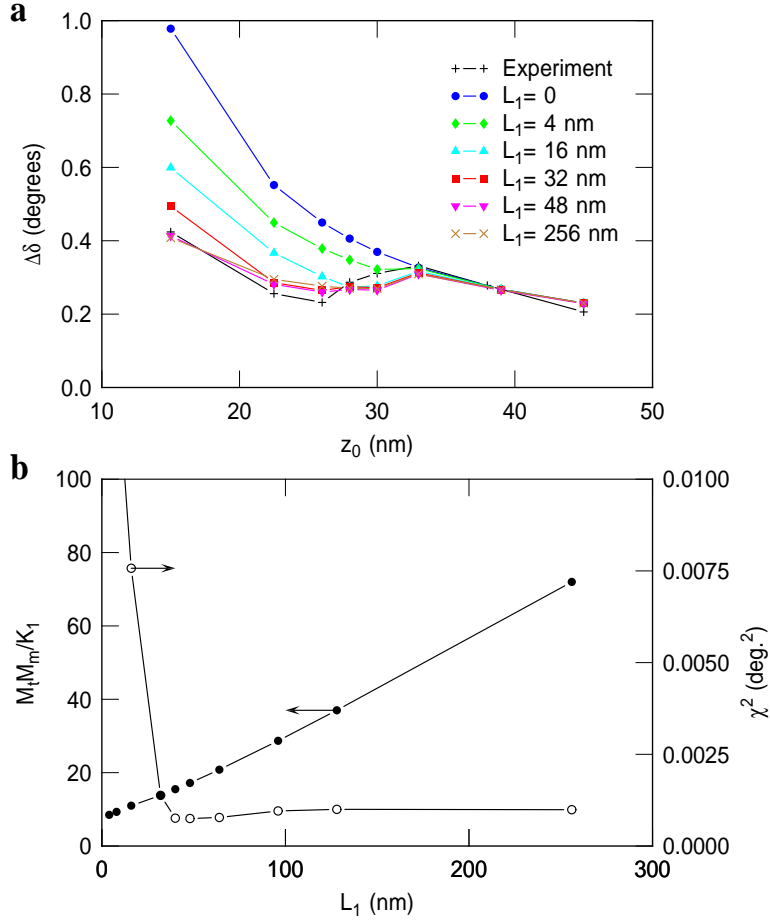}
\caption{\label{fig:zhiexcmd}
(color online)
(a) Full-scale variation in phase angle along cross-sections through the centers of the recorded tracks in
Figure \ref{fig:natnt2}b. The $+$ symbols represent experiment. The other symbols represent modeling
as described in the text. The modeling curves are labeled by the length $L_1$
of tip that switches magnetic moment orientation.
(b) Best fit values for $M_tM_m/K_1$, and $\chi^2 = \sum (\Delta \delta _{exp} - \Delta \delta _{mod})^2/(N-1)$
as a function of the tip switching length $L_1$, with $\sigma_w/\mu_0 L_1 K_1$ = 0. }
\end{figure}

Figure \ref{fig:zhiexcmd} compares the maximum minus the minimum value for $\delta$ along a cross-section
through the center of the bits at a bit density of 300 kfci
as a function of tip height $z_0$. The modeling results in this Figure
are labeled by the length $L_1$ of tip that is allowed to reorient its magnetic moment.
There are three parameters in this analysis - a global
multiplicative factor, $M_tM_m/K_1$,
and the reduced domain wall energy
$\sigma_w/\mu_0 L_1 K_1$.
Figure \ref{fig:zhiexcmd}b plots the best fit values for
$M_tM_m/K_1$ and $\chi^2 = \sum (\Delta \delta _{experimental} - \Delta \delta _{model})^2/(N-1)$
(N the number of data points). In all cases
the best fit value for $\sigma_w/\mu_0 L_1 K_1$ is 0, and
the $\chi^2$ value at $\sigma_w/\mu_0L_1K_1 =1$ is approximately double that
when $\sigma_w/\mu_0 L_1 K_1 =0$. Increasing
the domain wall energy requires larger switching fields: the best fit value at $L_1$= 64 nm
for $M_tM_m/K_1$ increases from 20.8 to 28.6 when $\sigma_w/\mu_0 L_1 K_1$ increases from 0 to 1.
The domain wall energy for Co is reported to be
$\sigma_w = 25\pm3 J/m^2$\cite{hehn1996}. This leads to $\sigma_w/\mu_0L_1K_1 =0.85$, using
$L_1=48nm$ and $K_1$=0.25 $M_t^2$ (neglecting crystalline anisotropy),
with $M_t=1.4 \times 10^6 A/m$, so that our fits are consistent
with the calculated wall energy, if one allows for a doubling of the best $\chi^2$ value.
The tip end may be magnetically poorly coupled to the rest of the
tip because of an inhomogeneity or grain boundary: it
appears (Figure \ref{fig:tipandgeo} inset) to have granularity on the scale of a few tens
of nm and a kink about 50 nm from its end.
We have observed qualitatively similar spatial
frequency doubling using several nanotube tips, particularly in the smallest diameter tips,
where inhomogeneities and weak magnetic coupling are fundamentally more difficult to avoid.

The shift $\Delta \omega$ in the resonance frequency $\omega_0$ of the cantilever due to a force gradient
between the tip and sample $dF_{z}/dz_0$ is given by $\Delta(\omega)/\omega_0 = -(dF_{z}/dz_0)/2k$,
where $k$ is the spring constant of the cantilever. The phase shift $\delta$ of the cantilever
response is then given by
\begin{equation}
\tan \delta = \frac{1}{Q} \frac{1}{\omega_0^{'}/\omega - \omega/\omega_0^{'}},
\label{eq:resonance}
\end{equation}
where $Q$ is the quality factor, $\omega$ is the driving frequency
and $\omega_0^{'}$ is the perturbed resonance frequency of the cantilever.
At $z_0=15 nm$ our model predicts an excursion of $\Delta (dF/dz)/\mu_0 M_t M_m \approx 2 \times 10^{-9} m$.
Using a driving frequency at optimal sensitivity $\omega= \omega_0^{'}(1-1/\sqrt{8}Q)$, $k=2.8 N/m$,
Q=350, $\omega_0 = 168 kHz$, and estimating $M_t = 1.4 \times 10^6 A/m$ \, \cite{otani2000}
and $M_m = 2 \times 10^5 A/m$ \cite{ross2001},
this corresponds to an excursion in the phase shift $\Delta \delta = 0.7^{o}$,
in reasonable agreement with the experimental value of $\Delta \delta = 0.4^{o}$
given the uncertainties in the values for the saturation
magnetizations.
Using $M_t|<H_x>_c|/K_1 <\sim $  2 \cite{stoner1948}, the best fit value
$M_tM_m/K_1 \sim$ 17 (Figure \ref{fig:zhiexcmd})
implies
a critical field of approximately 2.4$\times 10^4$ A/m,
much smaller than the saturation magnetization of cobalt of
1.4$\times 10^6$ A/m, but comparable
to a switching field of 3.2$\times 10^4$ A/m reported
for 30 nm thick, 0.34 $\mu$m wide,
2.04 $\mu$m long ellipsoidal amorphous cobalt nanodots \cite{johnson2000}.
This reduction in switching field could result from
competition between the crystalline and shape
anisotropies \cite{otani2000} if, for example, the uniaxial crystalline anisotropy favors
moment alignment along the tip radial direction, while the shape anisotropy favors the axial direction.
The tip material could also have a complicated structure incorporating grain and domain boundaries,
reducing the anisotropy energy.
The highly non-uniform fields in our case could also play a role in reducing the switching field.

Although our analysis has been presented using a specific model for the
tip dynamics, the conclusion that the tip moment flips at relatively small
fields can be presented simply: The magnetic field at the tip must be less
than the magnetic field at the medium surface, which is given by the saturation
magnetization of the medium. The fields required to flip the tip are expected,
using for example the Stoner-Wohlfarth model \cite{stoner1948}, to be about the saturation
magnetization of the tip, which is for epitaxial cobalt much larger than the
saturation magnetization of the medium. The switching fields of our nanotube,
just as for amorphous nanodots, are much smaller than those of epitaxial Co
nanodots \cite{bonet1999,otani2000}, which are comparable to the saturation magnetization of cobalt.
It might be possible to avoid switching and the attendant spatial frequency
doubling by developing processes to epitaxially coat the nanotube or by using
single-crystal nanorods. However, such tips would also generate larger local
magnetic fields at the sample, increasing the possibility of changing the magnetic
state of the sample, particularly for the smallest samples. Our results show that
reliable information on the moment orientations of the media can be inferred even
in the presence of tip switching if it is realised that the MFM signal is proportional
to the absolute magnitude of the tip-sample force gradient.

\ack{
We would like to thank Dennis Adderton of First Nano and Dr. Steve Minne of Veeco Instruments
for the AFM probes, and Dr. David Guarisco of Maxtor
Corporation for the recorded disks.
This work was supported
by the Center for Probing the Nanoscale (CPN), an NSF NSEC,
NSF Grant No. PHY-0425897, by NSF Grant No. DMR 0103548,
by the Dutch Foundation for Research on Matter (FOM),
the Netherlands Organization for Scientific Research (NWO),
and the Dutch STW NanoNed program.
}


\section*{References}
\begin{thebibliography}{99}

\bibitem{deng2004} Deng Z, Yenilmez E, Leu J, Hoffman J E, Straver E W J, Dai H, and Moler K A
2004 {\it Appl. Phys. Lett.} {\bf 85} 6263

\bibitem{deng2006} Deng Z, Yenilmez E, Reilein A, Leu J, Dai H and Moler K A
2006 {\it Appl. Phys. Lett.} {\bf 88} 023119

\bibitem{yenilmez2002}  Yenilmez E, Wang Q, Chen R J, Wang D W, and Dai H J
2002 {\it Appl. Phys. Lett.} {\bf 80} 2225

\bibitem{steifel1998} Steifel B, ``Magnetic Force Microscopy at Low Temperatures
and in Ultra High Vacuum - Application on High Temperature Superconductors",
Inauguraldissertation, University of Basel, 1998.

\bibitem{stoner1948} Stoner E C and Wohlfarth E P
1948 {\it Phil. Trans. Royal Soc. London} {\bf A 240} 599

\bibitem{kittel1946} Kittel C,
1946 {\it Physical Review} {\bf 70} 965

\bibitem{bonet1999} Bonet E, Wernsdorfer W, Barbara B, Benoit A, Mailly D and Thiaville A,
1999 {\it Phys. Rev. Lett.} {\bf 83} 4188

\bibitem{hehn1996} Hehn M, Padovani S, Ounadjela K and Bucher J P,
1996 {\it Phys. Rev. B} {\bf 54} 3428

\bibitem{otani2000} Otani Y, Kohda T, Novosad V, Fukamichi K, Yuasa S and Katayama T,
2000 {\it J. Appl. Phys.} {\bf 87} 5621

\bibitem{ross2001} Ross C A
2001 {\it Annu. Rev. Mater. Res.} 203

\bibitem{johnson2000} Johnson J A, Grimsditch M, Metlushko V, Vavassori P, Illic B,
Neuzil P and Kumar R
2000 {\it Appl. Phys. Lett.} {\bf 77} 4410

\end{thebibliography}
\end{document}